# Gamma relaxation in bulk metallic glasses

Stefan Küchemann and Robert Maaß*

*Department of Materials Science and Engineering, University of Illinois at Urbana-Champaign, 1304 West Green Street, Urbana 61801 IL, USA*

Studying the primary $\alpha$- and secondary $\beta$-relaxation process has contributed significantly to the understanding of the structure and rheology of metallic glasses. In this letter, we report on a third relaxation mechanism indicated by a maximum in the loss modulus at low temperatures, which we term $\gamma$-relaxation. Contrary to the $\alpha$- and $\beta$-relaxation mechanisms, this irreversible, low energy excitation causes a macroscopic rejuvenation, which we assign to non-affine atomic rearrangements in the matrix that are driven by thermal stresses during cooling. Observed in three different glassy alloys, the low temperature relaxation is identified as a general process in metallic glasses.

* electronic mail: rmaass@illinois.edu

It has been long recognized that exciting a structural material with a small oscillatory stimulus may give rich insights into the details of fundamental relaxation mechanisms. Examples of such are momentary relaxation of dielectrics [1], internal friction measurements of dislocations in crystals [2], or viscoelastic properties of polymers [3], colloids [4], or metallic glasses [5]. In particular, the case of dynamical mechanical spectroscopy, DMA, has been a versatile tool to study different kinds of structural excitations in disordered systems, where distinct peaks in the loss modulus as a function of temperature reveal one, two or three structural relaxation processes when conducted at a given frequency.

The analysis of relaxation mechanisms has played an essential role in the understanding of glasses as they reveal fundamental structural differences to crystals [6, 7]. In general, glassy systems universally exhibit two relaxations modes: the primary $\alpha$- and the secondary $\beta$-process [8, 9]. Primary excitations near the glass transition temperature, $T_g$, are associated with large cooperative atomic mobility leading to irreversible viscous flow. The secondary $\beta$-relaxation process below $T_g$ is understood as a cooperative rearrangement, but it is energetically reversible and reflects structural transitions on a much smaller energy scale [10-12]. In the particular case of metallic glasses, $\beta$-relaxations, also termed Johari-Goldstein relaxations [13], have been linked to reversible local transitions of small chain-like groups of atoms ([14, 15]) that collectively are mediating $\alpha$-transitions. Recently, evidence for a relaxation process observed at low temperatures (0.45$T_g$) has been reported for a very particular La-based glass formers, which was sought to find its origin in the specific chemistry of the system [16, 17].

Here we provide strong evidence for the existence of a general third structural relaxation mechanism in metallic glasses, which we observe in three glass forming alloys at around 0.2$T_g$-0.3$T_g$. This low temperature relaxation is very akin to the long known $\gamma$-relaxation in amorphous polymers, where it has been reported, for instance, in co-blockpolymers [18], epoxy resins [19, 20] and substituted polystyrenes [21].

We find that $\gamma$-relaxation in the here studied bulk metallic glasses represents an irreversible structural excitation, causing a remarkable enthalpy storage that is recovered during reheating below the glass transition temperature. The barrier energy of the $\gamma$-relaxation was found to be of the order of a few tenth eV, being distinctly separated from the typical barrier energy scale of the $\alpha$- and $\beta$-process.

Three different metallic glasses ($Zr_{58.5}Cu_{15.6}Ni_{12.8}Al_{10.3}Nb_{2.8}$ (Vit106a, Liquidmetal Technologies), $Zr_{66.5}Cu_{33.5}$, $Pd_{77.5}Cu_6Si_{16.5}$) were prepared in ribbon form with a thickness of ca. 40-110 μm. X-ray measurements were conducted to verify a glassy structure. In the following we focus on the results for Vit106a, and data for the other two alloys can be found in the supplementary material (SM). Mechanical spectroscopy measurements were performed using a TA Q800 DMA in tensile geometry. The measurements were carried out in air and liquid nitrogen was used as a cooling agent. All samples were preloaded with a static stress below 13 MPa (which corresponds to 1% of the yield stress) and periodically excited either with a controlled strain of 0.03% or in a force controlled mode with a dynamic stress below 13 MPa. Differential scanning calorimetry (DSC) measurements were performed using a Perkin Elmer DSC 8000. After equilibration at 323 K, the samples were heated to 753 K in the supercooled liquid regime ($T_g$= 674 K, $T_x$= 782 K) with 40 K/min. After cooling with 80 K/min down to 323 K, where the temperature was held for 1 min, the samples were heated again with 40 K/min to 753 K.

Figure 1 shows the loss modulus $E''$ as a function of temperature obtained by DMA measurements. Starting from room temperature the system was cooled to 116 K with a rate of 2 K/min while it was dynamically excited with a frequency of 5 Hz. Additionally, the sample was heated to 713 K above the glass transition temperature $T_g$ with the same heating rate and dynamic excitation frequency. In the temperature regime between room temperature and $T_g$, the typical $\alpha$-peak with the extensively investigated excess wing ($\beta$-relaxations) on the low-temperature side is seen. Conceptually visualized in a one-dimensional potential energy landscape (PEL), the $\alpha$-transition is said to reflect transitions between meta-basins, where shear deformation can lower the effective transition barriers, therefore triggering the system to attain another nearby state. This viscous flow means configurational changes of all atoms in the system and requires a certain amount of thermal energy. At lower temperatures, the preceding local string-like excitation involves a significantly smaller maximum ($\beta$-transitions) that is partially merged with the $\alpha$-peak and thus appears for the now discussed Zr-based alloy only as a low-temperature wing of the main $\alpha$-peak.

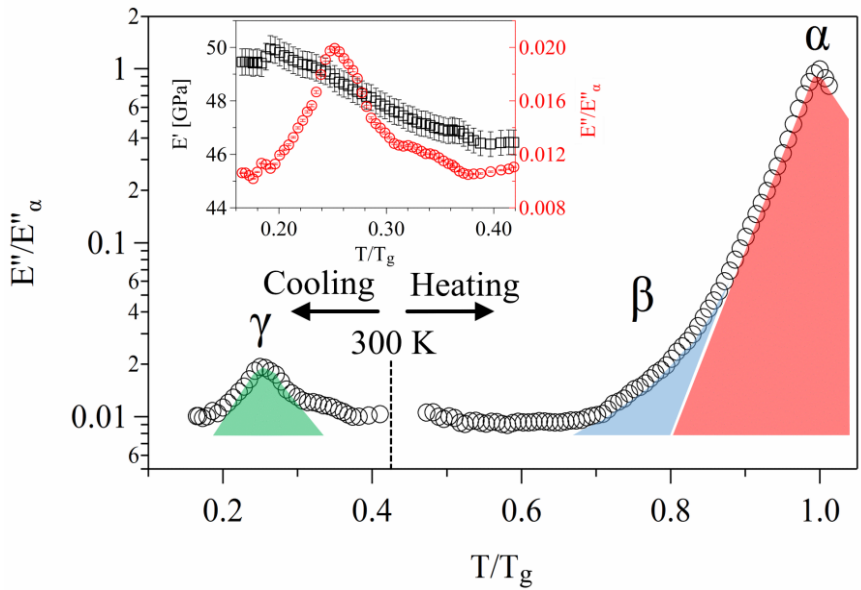

Figure 1: Relaxation spectrum for the bulk metallic glass Vit106a between 0.15×$T_g$ and $T_g$. Besides the well know $\alpha$- and $\beta$-process, a third relaxation process occurs around 0.26×$T_g$ ($\gamma$-peak). The inset shows a magnification of the peak in the loss modulus $E''$ at low temperatures and the corresponding storage modulus $E'$. The error bars in the loss and storage modulus in the inset result from machine specific confidence intervals.

Probing structural excitations upon cooling down from room temperature reveals a third distinct peak in the relaxation spectrum in Figure 1. At around 177 K ($0.26 \times T_g$) we observe a maximum in $E''$, indicating that the structural relaxation dynamics of metallic glasses is far more complex than hereto believed. In addition to Vit106a, we observe the same behaviour in $Zr_{66.5}Cu_{33.5}$ and $Pd_{77.5}Cu_6Si_{16.5}$ (see SM, Figure S1) with a third maximum in the mechanical loss spectrum at $0.26 \times T_g$ and $0.29 \times T_g$, respectively. This suggests that the third relaxation process is a general phenomenon in metallic glasses. In analogy to the low-temperature relaxation phenomenon of glassy polymers we term the low temperature mode $\gamma$-relaxation. The inset in Figure 1 shows a zoom-in of the low-temperature peak in $E''/E''_\alpha$ and the storage modulus upon cooling. For the temperature range of the $\gamma$-peak, the storage modulus has an overall slope of $dE/dT = 26.4 \pm 0.4$ MPa/K, which is in good agreement with the typically reported temperature dependence of the Young's modulus [22, 23].

To further quantify the nature and behaviour of this fast relaxation phenomena in metallic glasses, we conduct both a cooling-heating cycle, as well as two subsequent cooling cycles using the same cooling and heating rate and dynamical excitation frequency. The difference between both protocols is that in the latter mechanical excitation was only applied during cooling, whereas the sinusoidal dynamic strain amplitude was applied during cooling and heating in the former. It is found that the maximum in the loss modulus also occurs during reheating at 208 K ($0.31 \times T_g$), i.e. that there is a temperature difference of $\Delta T=52$ K between the $\gamma$-peak position during cooling and during reheating (Figure 2a), while in two subsequent cooling cycles, the $\gamma$-peak shifts from 174 K to 164 K (Figure 2b). It is also noticed in Figure 2 that the main loss peak exhibits a high temperature shoulder, which suggests an additional underlying process partially merged with the main relaxation mode. At this point we cannot distinctly separate between these two parts of the relaxation spectrum and treat both processes as a single relaxation mode.

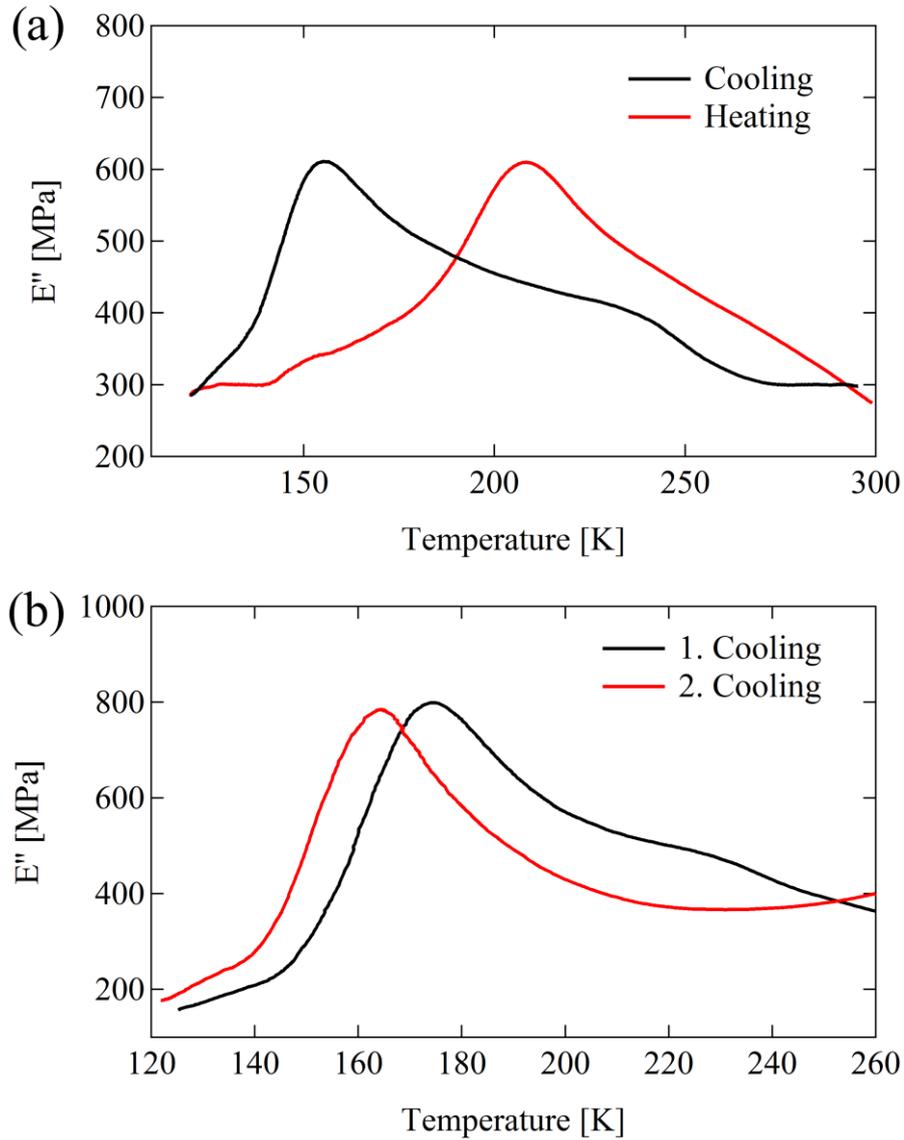

Figure 2: Cyclic behaviour of the γ-peak during cooling with 5 K/min and 7 Hz. (a) The cooling is followed by reheating while the sample is still dynamically excited. (b) Two successive cooling runs with the same sample. During the intermediate heating the sample was not mechanically excited.

Probing the γ-relaxation as a function of heating rate and excitation frequency allows accessing the barrier energy of the associated structural transitions. Clearly, the onset and the peak position shift to lower temperatures upon decreasing the frequency from 100 Hz to 1 Hz at a constant cooling rate of 2 K/min (Figure 3a). Evaluating the dependence of the γ-peak position on the frequency allows to derive the activation energy using the Kissinger method, as is shown as an inset in Figure 3b. Irrespective of choosing the peak position or the onset of the peak, we find an Arrhenius behaviour with a slope that yields an activation energy of 0.32 ±0.04 eV. Using the cooling rate $\Phi$, an activation energy of 0.22 ±0.08 eV was found (see SM, Figure S3c).

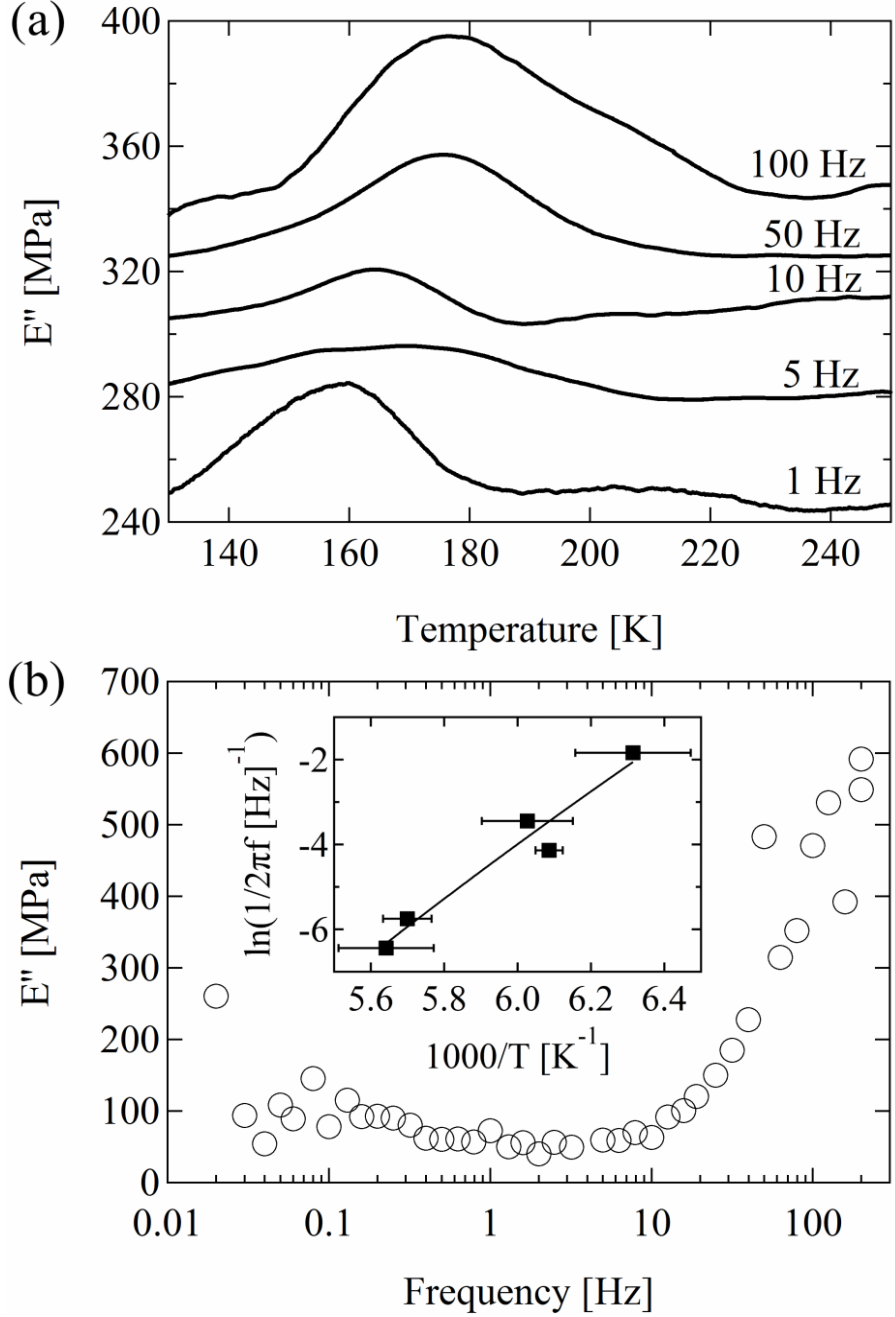

Figure 3: (a) Frequency dependence of the γ-process at a cooling rate of 2 K/min. See SM, Figure S3c, for the corresponding cooling-rate dependence. (b) Frequency sweep at 173 K from low to high frequencies. The inset in (b) shows the Arrhenius plot for the determination of the activation energy of the γ-process using the data of (a).

These values are distinctly lower than for β- (0.6-1.5 eV [10, 24]) or α-relaxations (>5.0 eV [25]). Therefore the γ-relaxation in metallic glasses is a fast process involving transitions across barriers that are at least two times smaller than for the primary and secondary relaxation mechanisms. Investigating the frequency dependence of the γ-relaxation with a sweep from 0.02 Hz to 200 Hz at a temperature of 173 K, reveals that $E''$ remains fairly constant in the lower frequency range, but there is a strong increase in the range from 10 Hz to 200 Hz (Figure 3b). Here, the upper and lower frequency bounds are limited by the machine. This increase in $E''$ at higher frequencies is due a gradual shift of the $\gamma$-peak to higher temperatures: at each frequency the maximum is effectively at a different temperature whilst the frequency scan is performed at a constant temperature. Therefore, the data in Figure 3b reflects a partial excitation of the relaxation dynamics at each frequency, with approximately the maximum being probed at the highest frequencies. In order to quantify the structural modifications due to the transitions marked by the $\gamma$-mode, we performed DSC measurements on specimen excited at different frequencies but cooled to the same minimum temperature. This protocol is expected to tune the structural state by exciting different fractions of the $\gamma$-peak. To this end, four samples were cooled with 10 K/min and dynamically excited with frequencies of 1 Hz, 5 Hz, 10 Hz and 100 Hz as indicated in Figure 4a. At a frequency of 1 Hz, the loss modulus as a function of temperature does not show a maximum in the studied temperature window, while at 5 Hz and 10 Hz there is a noticeable increase close to the low temperature limit of the investigated range. Only at the highest frequency of 100 Hz the entire maximum of the $\gamma$-relaxation is visible in the investigated temperature regime. A similar picture emerges upon studying the cooling-rate dependence of the $\gamma$-peak (see Figure S3 in the SM). Thus, by varying the excitation frequency we were able to tune the system state of the metallic glass. The DSC data for the samples of Figure 4a is displayed in Figure 4b. The difference in the heat flow between the first and second heating run was used to calculate the specific capacity difference $\Delta c_P$. For the analysis of the excess enthalpy which relaxes below $T_g$, we analysed the enclosed area between $\Delta c_P$ of the as-cast sample and $\Delta c_P$ of the samples which experienced $\gamma$-relaxation. As shown in Figure 4b, there is a remarkable increase in enthalpy release for the material that has undergone $\gamma$-relaxation. It is seen that relaxation due to heating begins as low as $0.5 \times T_g$. The enthalpy release in comparison to the enthalpy release of the as-cast sample below $T_g$ in Figure 4b amounts to $\Delta H_{1Hz} = 0.5$ kJ/mol, $\Delta H_{5Hz} = 0.7$ kJ/mol and $\Delta H_{10Hz} = 2.0$ kJ/mol and $\Delta H_{100Hz} = 5.1$ kJ/mol.

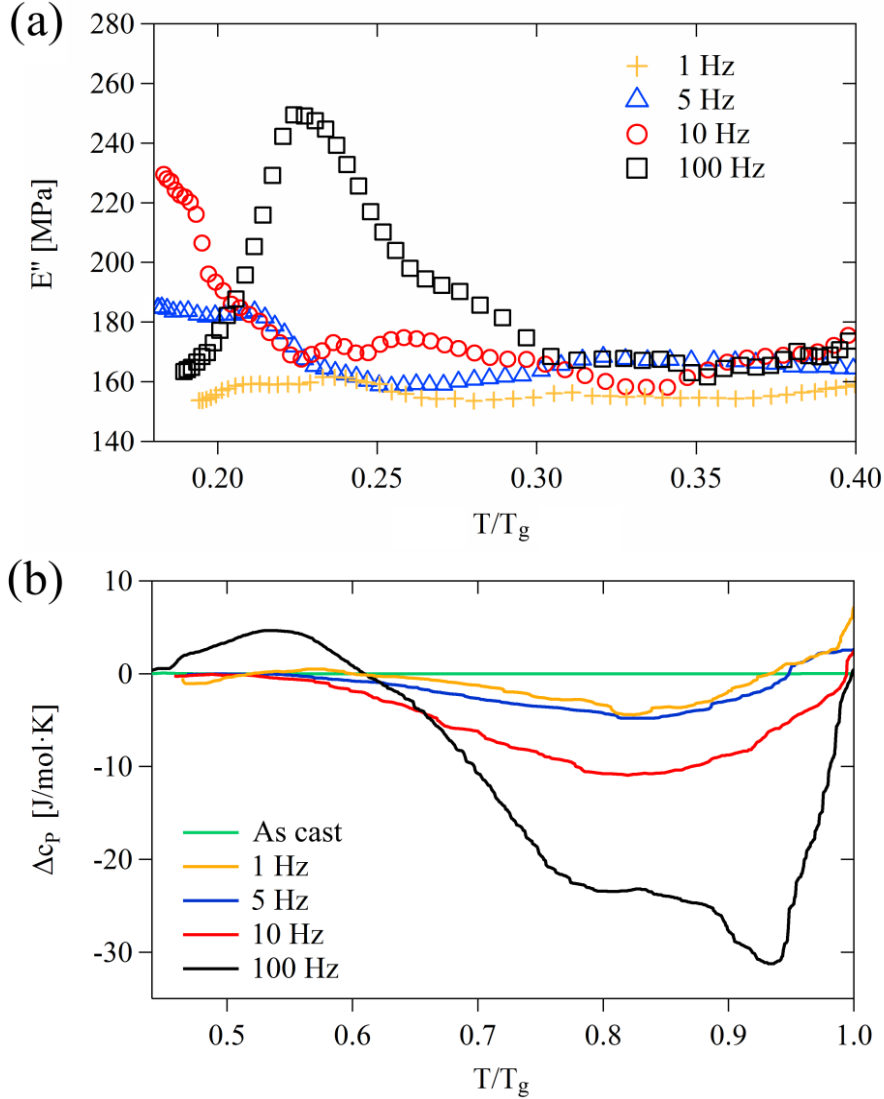

Figure 4: Characterisation of the system state due to the γ-relaxation. (a) Four different metallic glass samples were cooled with 10 K/min and excited with frequencies mentioned in the legend. (b) Specific heat capacity difference as a function of the normalized temperature for as cast glass state and four samples which experienced different fraction of the γ-relaxation from panel (a).

Given the clearly rejuvenated structure after cyclic excitation during cooling, the terminology "relaxation" may in fact be seen as a misnomer if placed into the concept of high temperature DMA experiments that indeed lead to a reduction in enthalpy; that is aging. The here found data conveys a picture in which the cooled metallic glass undergoes structural transitions that increase the energy state of the system. The stored enthalpy of 5.1 kJ/mol found here (Figure 4, 100 Hz) corresponds to $0.8 \times \Delta H_x$ of the heat of crystallization and to $0.59 \times \Delta H_f$ of the heat of fusion [26]. We attribute the difference in energy storage in comparison to cryogenic cycling to the small oscillatory excitation that will facilitate atomic configurational changes and thus amplify the enthalpy storage. Similarly, the excess enthalpy stored in the metallic glass due to cryogenic cycling was found to saturate after a few cycles only [27], being very compatible with the fact that the local structural transition defining the *γ*-relaxation shifts to lower temperatures with decreasing frequency, which eventually leads to saturation in a given temperature window. We thus conclude that *γ*-relaxation is driven by local non-affine thermal stresses that develop during cooling, which is distinctly different to the observed *β*-mode that primarily signifies a thermally driven relaxation (aging) process during heating.

In Figure 2b we observe a shift of the peak position to lower temperatures during a second cooling run in comparison to the first run. In order to interpret this shift we consider that the *γ*-relaxation causes a small configurational change. Once the system returned to room temperature, parts of this structural modification remains and can be recovered as stored enthalpy in a DSC scan. During the second cooling cycle, a structure different to the first cycle is probed. Now the *γ*-peak shifts to lower temperatures, because the major part of the structure has already seen thermal stresses due to a heterogeneous contraction during the first cooling cycle. This means, we understand the shift to lower temperatures during the second cooling cycle as a signature of higher thermal stresses needed to activate the rejuvenation manifested by the *γ*-peak during the first cooling cycle.

When applying an external mechanical stimulus during reheating, the mechanical-loss peak occurs at significantly higher temperatures in comparison to the one during cooling, with a temperature difference of $\Delta T = 52$ K. At such high temperatures, we do not expect that the system undergoes an additional rejuvenation process since the heterogeneous stresses due to the thermal bias are reduced in comparison to temperatures where the γ-peak occurs during cooling. Therefore, we believe that the peak observed during heating must rather be an indication for a thermally induced relaxation of the structure previously excited by heterogeneous thermal stresses during cooling.

The primary and secondary relaxation process are often qualitatively discussed and described in the picture of a PEL, where transitions between two metabasins (*α*-process) and transition between sub-basins within a single metabasin (*β*-process) co-exist. The general consensus is that one *α*-transition requires a number of small sub-basin transitions. In the view of the small activation energy found here for the *γ*-process, a modification of this picture is required, including an even finer barrier energy structure. Whether there exists, as for primary and secondary relaxation, a relationship between *γ*- and *β*-transition remains to be explored by future studies.

The small activation energy for *γ*-transition also allows to conclude on the size of the event. From simulations and measurements of the activation volume it is estimated that the large scale correlated motion during the *α*-process includes 140-660 atoms [28-30], while there are about 10-50 atoms involved in the local string like rearrangement during the *β*-process [14, 31, 32]. When relating the activation energies to these quantities, the *γ*-relaxation could only be associated with a small local configurational change of a few atoms. This is very similar to *γ*-relaxation in amorphous polymers. In these systems the *γ*-relaxation could be linked to the motion of side-groups, which only contain a few atoms [20, 33] and that are likely to change their configurations due to local bond breaking of hydrogen or van der Waals bonds [34]. Clearly, the atomic mechanisms underlying $\gamma$-relaxation in polymers is different than in metallic glasses, but the very local bond loosening mechanism in polymers is compatible with our interpretation of the $\gamma$-relaxation in metallic glasses being a plastic structural process involving atomic rearrangements that are smaller than for the $\beta$-mode and therefore likely below an interatomic distance. We finally note that recent atomistic simulations that probe thermally activated transitions at low temperatures (~$0.17T_g$) indeed have identified cooperative activity, where the typical displacements are smaller than an atomic radius [35]. We finally note that a fast relaxation process at higher temperature was recently reported by Qiao et al. [36], who observed an unexpected decoupling of the relaxation mechanism below the glass transition temperature into a fast stress driven and a slow thermally activated mode. The activation energy of those modes, however, differs significantly from the one of the $\gamma$-process and therefore probably mark a different underlying mechanism.

In summary, we reveal a general low temperature relaxation mechanism seen between $0.26T_g$ and $0.29T_g$ for three different glass forming alloys ($Zr_{58.5}Cu_{15.6}Ni_{12.8}Al_{10.3}Nb_{2.8}$, $Zr_{66.5}Cu_{33.5}$, $Pd_{77.5}Cu_6Si_{16.5}$) during cooling. Similar to the well-known primary $\alpha$-process and the secondary $\beta$-process, this low-temperature relaxation exhibits a maximum in the loss modulus. Here we name the observed relaxation mechanism $\gamma$-relaxation based on its resemblance to the low temperature loss mechanism in amorphous polymers. Our data reveals that $\gamma$-relaxation presents a fast structural transition process, with a barrier energy that is distinctly lower than for $\beta$-transitions. Due to the pronounced rejuvenation driven by $\gamma$-transitions, we argue that such relaxation events are originating from atomic-scale thermal stresses that develop during cooling. With this additional relaxation process, we anticipate that our findings form an important addition to the picture of the potential energy landscape, atomistic relaxation mechanisms in disordered materials, and that the detailed characteristics of this low temperature relaxation mode will shed new light onto the understanding of rejuvenation and energy storage processes in metallic glasses in general.

## SUPPLEMENTARY MATERIAL

See supplementary material for additional DMA measurements on $Zr_{66.5}Cu_{33.5}$ and $Pd_{77.5}Cu_6Si_{16.5}$, frequency, and cooling rate dependencies.

## ACKNOWLEDGMENTS

The authors would like to thank R. Espinosa-Marcal for providing a DSC and J. Lopez for supporting the calorimetry measurements. S.K. and R.M. are grateful for fruitful discussions with A.L. Greer and P.M. Derlet. Parts of the work have been carried out at the Frederick Seitz Materials Research Laboratory of the University of Illinois at Urbana-Champaign, and technical support by K. Walsh is acknowledged. R.M. thanks UIUC and the Department of Materials Science and Engineering for start-up funding.

[1] P. Lunkenheimer, A. Loidl, Chemical Physics 284(1) (2002) 205-219.
[2] A. Seeger, Philosophical Magazine 1(7) (1956) 651-662.
[3] J. Mijovic, H. Lee, J. Kenny, J. Mays, Macromolecules 39(6) (2006) 2172-2182.
[4] P. Segre, S. Meeker, P. Pusey, W. Poon, Physical review letters 75(5) (1995) 958.
[5] J. Qiao, J. Pelletier, Journal of Materials Science & Technology 30(6) (2014) 523-545.
[6] R. Kohlrausch, Annalen der Physik 167(2) (1854) 179-214.
[7] K. Ngai, Relaxation and diffusion in complex systems, Springer Science & Business Media2011.
[8] J. Hachenberg, K. Samwer, Journal of non-crystalline solids 352(42) (2006) 5110-5113.
[9] W. Gotze, L. Sjogren, Reports on progress in Physics 55(3) (1992) 241.
[10] H.B. Yu, W.H. Wang, H.Y. Bai, K. Samwer, National Science Review 1(3) (2014) 429-461.
[11] H.-B. Yu, W.-H. Wang, K. Samwer, Materials Today 16(5) (2013) 183-191.
[12] J.S. Harmon, M.D. Demetriou, W.L. Johnson, K. Samwer, Physical Review Letters 99(13) (2007).
[13] Goldstein. M, J. Chem. Phys. 51(9) (1969) 3728-&.
[14] S. Swayamjyoti, J.F. Löffler, P.M. Derlet, Physical Review B 89(22) (2014) 224201.
[15] H. Teichler, Journal of non-crystalline solids 293 (2001) 339-344.
[16] Q. Wang, S. Zhang, Y. Yang, Y. Dong, C. Liu, J. Lu, Nature communications 6 (2015).
[17] L.Z. Zhao, R.J. Xue, Z.G. Zhu, K.L. Ngai, W.H. Wang, H.Y. Bai, The Journal of Chemical Physics 144(20) (2016) 204507.
[18] M. Schwabe, R. Rotzoll, S. Küchemann, K. Nadimpalli, P. Vana, K. Samwer, Macromolecular Chemistry and Physics 211(15) (2010) 1673-1677.
[19] D.E. Kline, Journal of Polymer Science 47(0149) (1960) 237-249.
[20] G.A. Pogany, Polymer 11(2) (1970) 66-&.
[21] M. Baccaredda, E. Butta, V. Frosini, S. Depetris, Materials Science and Engineering 3(3) (1968) 157-+.
[22] A. Kawashima, Y. Yokoyama, I. Seki, H. Kurishita, M. Fukuhara, H. Kimura, A. Inoue, Mater. Trans. 50(11) (2009) 2685-2690.
[23] A. Kawashima, Y. Zeng, M. Fukuhara, H. Kurishita, N. Nishiyama, H. Miki, A. Inoue, Materials Science and Engineering A 498(1-2) (2008) 475-481.
[24] W.H. Wang, Journal of Applied Physics 110(5) (2011) 053521.
[25] J. Qiao, R. Casalini, J.-M. Pelletier, H. Kato, The Journal of Physical Chemistry B 118(13) (2014) 3720-3730.
[26] I. Gallino, M.B. Shah, R. Busch, Acta materialia 55(4) (2007) 1367-1376.
[27] S. Ketov, Y. Sun, S. Nachum, Z. Lu, A. Checchi, A. Beraldin, H. Bai, W. Wang, D. Louzguine-Luzgin, M. Carpenter, Nature 524(7564) (2015) 200-203.
[28] S. Mayr, Physical review letters 97(19) (2006) 195501.
[29] D. Pan, A. Inoue, T. Sakurai, M. Chen, Proceedings of the National Academy of Sciences 105(39) (2008) 14769-14772.
[30] M. Schwabe, S. Küchemann, H. Wagner, D. Bedorf, K. Samwer, Journal of Non-Crystalline Solids 357(2) (2011) 490-493.
[31] Y. Fan, T. Iwashita, T. Egami, Nat Commun 5 (2014) 5083.
[32] T.C. Hufnagel, C.A. Schuh, M.L. Falk, Acta Materialia 109 (2016) 375-393.
[33] E.A.W. Hoff, D.W. Robinson, A.H. Willbourn, Journal of Polymer Science 18(88) (1955) 161-176.
[34] R.D. Andrews, T.J. Hammack, Journal of Polymer Science Part B-Polymer Letters 3(8PB) (1965) 659-&.
[35] P.M. Derlet, unpublished results (2016).
[36] J. Qiao, Y.-J. Wang, L. Zhao, L. Dai, D. Crespo, J. Pelletier, L. Keer, Y. Yao, Physical Review B 94(10) (2016) 104203.
[37] J. Hachenberg, D. Bedorf, K. Samwer, R. Richert, A. Kahl, M.D. Demetriou, W.L. Johnson, Applied Physics Letters 92(13) (2008) Art. No. 131911.

# Supplementary material

## Gamma relaxation in bulk metallic glasses

Stefan Küchemann and Robert Maaß*

*Department of Materials Science and Engineering, University of Illinois at Urbana-Champaign, 1304 West Green Street, Urbana 61801 IL, USA*

**Details of the sample treatment prior to the measurements**

The metallic glass sheets made of Vit106a were purchased from Liquidmetal Technologies and are labeled as "as-cast" since the samples did not experience any treatment prior to the shown measurements in our laboratory. Liquidmetal Technologies applies a short annealing below the glass transition temperature of a few seconds. No other treatments than the DMA measurements were applied prior to the DSC measurements. Furthermore the DSC curve of the as cast sample (Figure 4b in the main manuscript) also shows the absence of any sub-$T_G$ relaxation indicating that the internal stresses introduced by the quenching have been removed in this sub-$T_G$ annealing performed by Liquidmetal Technologies.

**Gamma-relaxation in other bulk metallic glass alloys**

In addition to the mechanical spectroscopy on Vit106a reported in the main manuscript, we show here that the $\gamma$-relaxation occurs in stoichiometrically different types of metallic glass formers. Figure S1a shows an additional relaxation peak in the mechanical loss spectrum of the binary metallic glass $Zr_{66.5}Cu_{33.5}$ that occurs at a temperature of $0.26T_g$ in addition to the high temperature $\alpha$- and $\beta$-relaxation. Figure S1b displays the mechanical loss as a function of temperature for the ternary $Pd_{77.5}Cu_6Si_{16.5}$, showing qualitatively the same behavior as the two Zr-based glass former with the $\gamma$-relaxation maximum at $0.29T_g$. The Pd-based alloy exhibits a slightly lower relative amplitude of the $\gamma$-peak in comparison to Zr-based glasses. Note that the $\beta$-relaxation of both $Zr_{66.5}Cu_{33.5}$ and $Pd_{77.5}Cu_6Si_{16.5}$ is not manifested as a separate peak in the relaxation behavior during heating, but rather as a wing of the main $\alpha$-relaxation peak. This is consistent with earlier observations for these systems[10]. Figure S1 indicates that the here identified $\gamma$-relaxation may be a universal feature of metallic glasses.

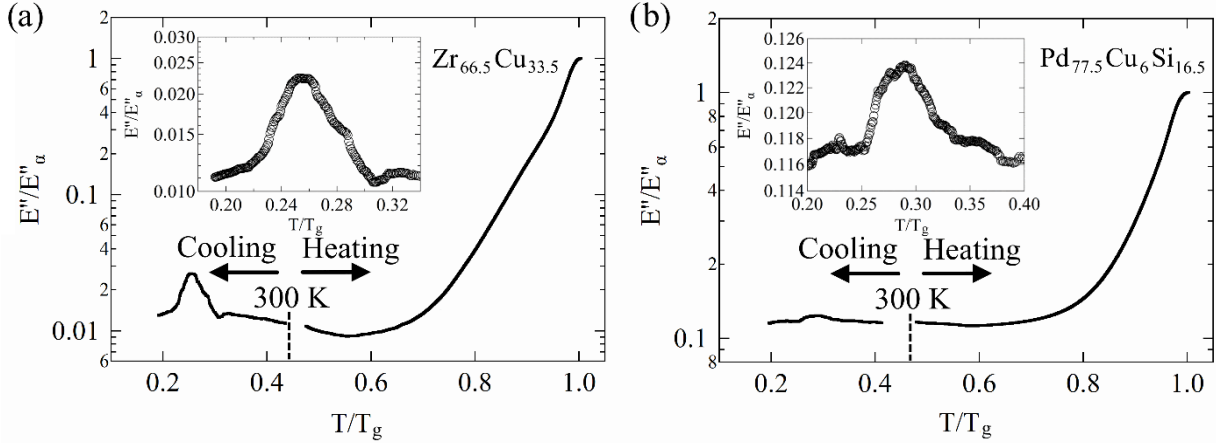

Figure S2: Normalized mechanical loss modulus as a function of temperature between $0.2T_g$ and $T_g$ of (a) binary $Zr_{66.5}Cu_{33.5}$ and (b) ternary $Pd_{77.5}Cu_6Si_{16.5}$. The insets highlight the low temperature γ-relaxation peak that occurs during cooling.

**Frequency dependence of the gamma-relaxation**

Here we provide additional data that exemplifies the determination of the peak position used to quantify the frequency dependence of the γ-relaxation displayed in the inset of Figure 3b of the main text. Figure S2 shows the mechanical loss E'' of Vit106a during cooling from room temperature down to 120 K in separate panels. The maximum results from standard Lorentzian fit of the peak (red line in Figure S2) given by $E''(T) = E_0'' + \frac{A}{(T-T_0)^2+B}$, with the amplitude A, the full width at half maximum B, the peak position $T_0$ and the constant offset $E_0''$.

In this simple form the Lorentzian fit does not account for any non-constant background, as for instance a slope or any asymmetry potentially arising due to the existence of a sub-peak. Therefore we restricted the fitting range to the upper part of the maximum and did not fit the background which is sufficient for the determination of the peak position. As discussed in the main manuscript, the maximum shifts to higher temperatures with increasing excitation frequency. Note that the measurement signal is affected by the size of the sample (see for instance top right panel in Figure S2 with an excitation frequency of 5 Hz). For samples with a smaller width the applied force was reduced so that the applied stress remains constant which causes the side effect of an increase in the noise.

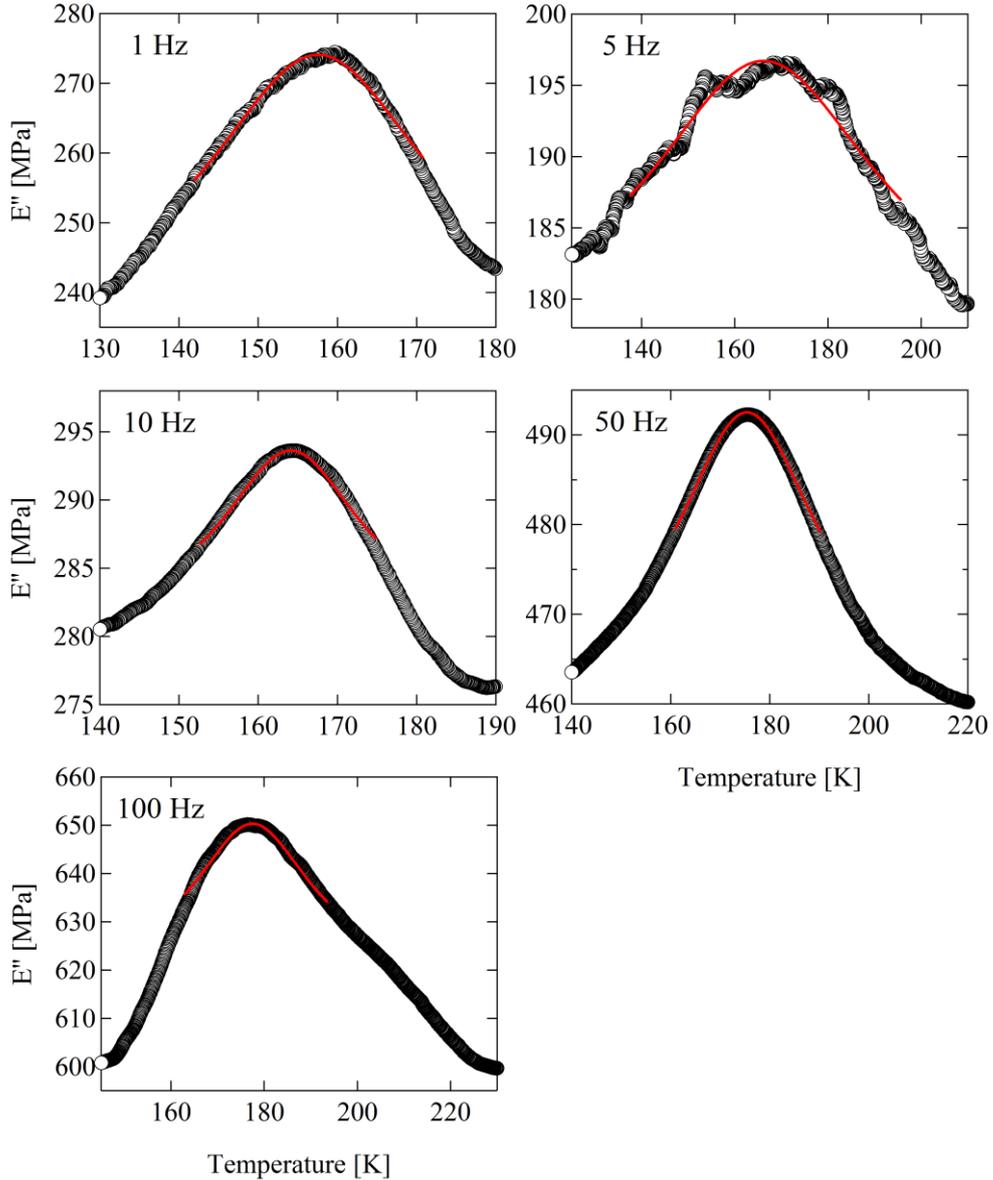

Figure S2: Mechanical loss of Vit106a during cooling as a function of temperature for five different frequencies. The red lines are Lorentzian fits to the peak.

**Cooling-rate dependence of the gamma-relaxation**

Additional to the frequency dependence of the $\gamma$-relaxation shown in the main manuscript in Figure 3a, the third relaxation mode also exhibits a dependence on the cooling rate. Figure S3a shows the mechanical loss as a function of temperature during for different cooling rates at 5 Hz of Vit106a. For the quantitative analysis of the peak shift as a function of cooling rate, the $\gamma$-relaxation peak was fitted with a Lorentzian fit and its onset during cooling is determined via the intersection of two tangents: one before the onset and one during the increase of the loss modulus at the high temperature side of the $\gamma$-relaxation (Figure S3a). The onset has been determined ten times independently with varying fitting ranges for the tangents in order to account for the underlying noise in the signal. Figure S3b shows the magnified graph of the measurement with a cooling rate of 0.5 K/min and a frequency of 5 Hz. Note again that the noise is enhanced due to lower absolute force levels in small samples. However the maximum and the onset can be well identified. Figure S3c shows an Arrhenius plot of the cooling rate $\Phi$ and the peak temperature of the $\gamma$-relaxation. The slope yields an activation energy of the $\gamma$-relaxation in Vit106a of $E_A = 0.22\pm0.08$ eV. This value is somewhat smaller than the activation energy of $E_A = 0.32\pm0.04$ eV obtained from the variation of the frequency at constant cooling rate reported in the main text (see Figure 3c in the main text). Such differences in barrier energy depending on the chosen experimental variable are typical for DMA measurements. For example, previous measurements clearly show that when determining the barrier energy for $\beta$-relaxations from a heating rate dependence, the obtained values can differ substantially from the barrier energies found when conducting a set of experiments where the mechanical frequency is varied [10, 37].

Figure S3a demonstrates that the $\gamma$-relaxation shifts to higher temperatures with decreasing cooling rate $\Phi$. This fact needs to be kept in mind when considering the stored enthalpy that was measured for samples excited at different frequencies and with $\Phi=10$ K/min (see Figure 4a and 4b in the main manuscript). In all of these measurements (Figure 4 and Figure S3), the lowest temperature (final temperature during cooling) was identical in each type of series as to unambiguously relate the $\gamma$-relaxation to the enthalpy storage in the glass, and to rule out any influence of the minimum temperature.

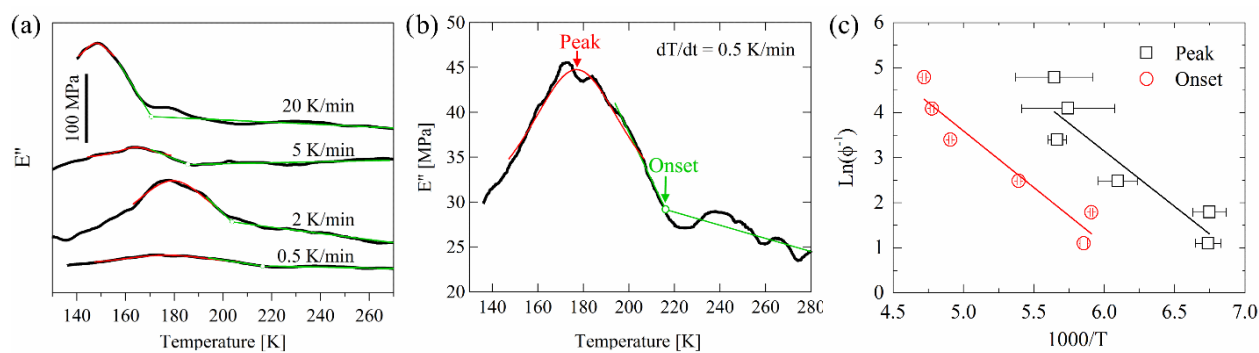

Figure S3: Dependence of the peak location on the excitation frequency at a constant frequency. (a) Mechanical loss measured at 5 Hz applying different cooling rates. The curves are shifted for clarity along the abscissa. The onsets have been determined via two intersecting tangents (green lines), the first one describing the base line and the second one the high temperature side of the loss peak. The red lines are Lorentzian fits for the determination of the peak position. (b) Magnification of the mechanical loss as a function of temperature at a cooling rate of 0.5 K/min. (c) Arrhenius plot of the cooling rate $\Phi$ and the peak temperature T determined via an asymmetric Lorentzian fit.

We finally note that there is a cooling-rate dependence on the enthalpy storage, as the DSC measurements of the material cooled with 2 K/min and excited at 10 Hz and 100 Hz (Figure 3a) yield an excess enthalpy release below the glass transition temperature relative to the as-cast material of 1.25 and 1.37 kJ/mol, respectively. This is lower than observed for a cooling rate of 10 K/min.